# Pseudocontact shifts
# from mobile spin labels

E. Suturina[*], Ilya Kuprov

*School of Chemistry, University of Southampton,*
*Highfield Campus, Southampton, SO17 1BJ, UK*

[*]Corresponding author: e.suturina@soton.ac.uk

**Abstract**

This paper presents a detailed analysis of the pseudocontact shift (PCS) field induced by a mobile spin label that is viewed as a probability density distribution with an associated effective magnetic susceptibility anisotropy. It is demonstrated that non-spherically-symmetric density can lead to significant deviations from the commonly used point dipole approximation for PCS. Analytical and numerical solutions are presented for the general partial differential equation that describes the non-point case. It is also demonstrated that it is possible, with some reasonable approximations, to reconstruct paramagnetic centre probability distributions from the experimental PCS data.

## 1. Introduction

Pseudocontact shift (PCS) is an additional contribution to the nuclear chemical shift caused by the presence of a paramagnetic centre in close proximity to the nucleus in question.[1,2] PCS is very well researched and is widely used as a source of structural restraints for paramagnetic metalloproteins.[3-7] Even when a protein is not naturally paramagnetic, the commonly occurring calcium, magnesium and zinc binding sites would usually coordinate a lanthanide well.[8] In combination with artificially introduced lanthanide-containing tags PCS is also used to determine relative orientations of protein domains.[9] In magnetic resonance imaging it is useful as a reporter for local pH and oxidation potential.[10,11]

General equations describing chemical shielding, obtained by an assiduously systematic application of perturbation theory, are due to Ramsey.[12,13] The first paper dealing with a point dipole approximation for chemical shift was published by McConnell[14] in 1957 – he noted that shielding by sufficiently distant electrons could be expressed *via* an effective magnetic susceptibility tensor; for paramagnetic molecules this tensor is a function of the spin Hamiltonian parameters of the paramagnetic centre.[15] Analytical treatments for specific classes of *d*- and *f*- transition metal complexes using ligand field theory have been reported by Bleaney,[16] Golding,[17] and Stiles.[18-21]

Modern quantum chemistry defines paramagnetic shift as the Frobenius inner product between the nuclear hyperfine coupling tensor and the magnetic susceptibility tensor.[22,23] Both parameters are difficult to compute because they have contributions from spin-orbit coupling and often require non-perturbative treatment of relativistic effects within multi-configurational *ab initio* methods,[24,25] as well as conformational averaging. Accurate quantum chemical calculations are therefore limited to a few dozen atoms.

Structural biologists couldn't care less – most nuclei in macromolecules are far enough away from the paramagnetic centre for McConnell's' approximation to be accurate[2,26] and the resulting formula for the pseudocontact shift produced by a point source with a given magnetic susceptibility anisotropy has been of great service to protein structure and dynamics research over the last 30 years.[8,27] Excellent software packages exist that make PCS analysis in proteins and nucleic acids straightforward and informative.[5,28]

There remains one important unsolved problem: pseudocontact shift prediction and analysis in large systems that feature fast conformational mobility of the paramagnetic centre. For such systems the point dipole approximation is no longer valid at short distances and quantum chemical calculations are prohibitively expensive. For this reason, PCS measurements in close proximity to the tag are often excluded from the analysis because they are not expected to fit the point dipole formula.[29]

In this communication we introduce an analytical approach based on the recently published partial differential equation for PCS[30] that views the paramagnetic centre as a probability density distribution in three dimensions. This approach clarifies the key features of that density that affect PCS. It may also be used to recover the distribution itself from the experimental PCS data.

## 2. Pseudocontact shift from a point paramagnetic centre

The point source formula for PCS is best derived using a classical physics argument. Placed in an external magnetic field $\mathbf{B}_0$, a paramagnetic centre would acquire the following magnetic dipole moment:

$$\boldsymbol{\mu}_e = \boldsymbol{\chi} \cdot \mathbf{B}_0 / \mu_0 \qquad (1)$$

where $\mu_0$ is the vacuum permeability and $\chi$ is the magnetic susceptibility tensor.[26] This linear response assumption is valid for an ensemble of non-interacting paramagnetic centres when $\mu_e^T \cdot B_0 \ll kT$ – true for most metalorganic systems at room temperature. For a point centre, the induced dipole creates the following magnetic field at the relative position $\mathbf{r}$:

$$\mathbf{B}_1 = \frac{3\mu_0}{4\pi r^3} \mathbf{D}(\hat{\mathbf{r}}) \cdot \boldsymbol{\mu}_e, \qquad (2)$$

where the dipolar matrix is:

$$\mathbf{D}(\hat{\mathbf{r}}) = \frac{1}{r^2} \begin{bmatrix} x^2 - \frac{r^2}{3} & xy & xz \\ xy & y^2 - \frac{r^2}{3} & yz \\ xz & yz & z^2 - \frac{r^2}{3} \end{bmatrix} \qquad (3)$$

It is easy to demonstrate that the dipolar matrix only depends on the direction of the position vector $\hat{\mathbf{r}} = \{\theta, \varphi\}$ and does not depend on its length $r = |\mathbf{r}|$ – this provides a clean separation of coordinates that will be useful below.

The change in the energy of a nuclear magnetic moment $\boldsymbol{\mu}_n$ produced by placing it at the position $\mathbf{r}$ relative to the paramagnetic centre would be

$$\Delta E = -\boldsymbol{\mu}_n^T \cdot \mathbf{B}_1 \qquad (4)$$

The associated chemical shift tensor is the second derivative of $\Delta E$ with respect to $\boldsymbol{\mu}_n$ and $\mathbf{B}_0$:[12,31]

$$\boldsymbol{\sigma}^{\text{point}} = -\frac{\partial^2 E}{\partial \boldsymbol{\mu}_n^T \partial \mathbf{B}_0} = \frac{3}{4\pi r^3} \mathbf{D}(\hat{\mathbf{r}}) \cdot \boldsymbol{\chi} \qquad (5)$$

The isotropic average of this tensor is the familiar point dipole expression for the pseudocontact shift:[14,26]

$$\sigma^{\text{point}} = \frac{1}{3}\text{Tr}\left[\boldsymbol{\sigma}^{\text{point}}\right] = \frac{1}{4\pi r^3} \text{Tr}\left[\mathbf{D}(\hat{\mathbf{r}}) \cdot \boldsymbol{\chi}\right] = \frac{1}{4\pi r^3}\left(\frac{\mathbf{r}^T \cdot \boldsymbol{\chi} \cdot \mathbf{r}}{\mathbf{r}^T \cdot \mathbf{r}} - \frac{1}{3}\text{Tr}(\boldsymbol{\chi})\right) \qquad (6)$$

that may also be rewritten *via* second-rank spherical harmonics:

$$\sigma^{\text{point}} = \frac{1}{4\pi r^3} \sum_{m=-2}^{2} \chi_m Y_2^m(\hat{\mathbf{r}}) \qquad (7)$$

where $\chi_m$ are the irreducible spherical components of $\boldsymbol{\chi}$:

$$\chi_{\pm 2} = +\sqrt{\frac{2\pi}{15}}\left(\chi_{XX} - \chi_{YY} \mp i(\chi_{XY} + \chi_{YX})\right)$$

$$\chi_{\pm 1} = -\sqrt{\frac{2\pi}{15}}\left(\chi_{XZ} + \chi_{ZX} \mp i(\chi_{YZ} + \chi_{ZY})\right) \qquad (8)$$

$$\chi_0 = +\sqrt{\frac{4\pi}{45}}\left(2\chi_{ZZ} - (\chi_{XX} + \chi_{YY})\right)$$

It is important to note that the isotropic part and the first spherical rank component of the magnetic susceptibility tensor do not enter the equation for PCS. The five irreducible spherical tensor parameters in Eq. (8)

may also be expressed as axiality and rhombicity, along with the three parameters (*e.g.* Euler angles) specifying the orientation of the principal axis frame.

## 3. Pseudocontact shift from a distributed paramagnetic centre

A less well explored situation is when the paramagnetic centre is distributed with some probability density $\rho(\mathbf{r}_e)$ within the molecular structure. In such a situation, the magnetic susceptibility tensor would also in general be position-dependent. Integration of the point PCS expression in Eq. (5) over the probability density produces the following expression for the effective dipolar shift tensor at position $\mathbf{r}$:

$$\boldsymbol{\sigma}(\mathbf{r}) = \frac{3}{4\pi} \int \frac{\mathbf{D}(\mathbf{r}_e - \mathbf{r})}{|\mathbf{r}_e - \mathbf{r}|^3} \boldsymbol{\chi}(\mathbf{r}_e) \rho(\mathbf{r}_e) d^3\mathbf{r}_e \tag{9}$$

This integral is a convolution of the dipolar matrix $\mathbf{D}$ divided by the cube of the distance with the product of susceptibility tensor and probability density:

$$\boldsymbol{\sigma}(\mathbf{r}) = \frac{3}{4\pi} \frac{\mathbf{D}(\mathbf{r})}{r^3} * [\boldsymbol{\chi}(\mathbf{r}) \rho(\mathbf{r})] \tag{10}$$

The simplest way to proceed is to use the Fourier transform because convolution is equivalent to multiplication in the *k*-space, and the Fourier transform of the dipolar matrix is very simple:

$$\mathrm{FT}_+\left[\frac{3}{4\pi r^3} \mathbf{D}(\hat{\mathbf{r}})\right] = -\mathbf{D}(\hat{\mathbf{k}}) \tag{11}$$

where $\hat{\mathbf{k}}$ is the angular part of the *k*-space vector $\mathbf{k}$. Another useful property of the dipolar matrix is that the inverse Fourier transform of its product with a function in *k*-space can be expressed as an action of the differential operator in real space:

$$\boldsymbol{\sigma}(\mathbf{r}) = -\mathrm{FT}_-\left[\mathbf{D}(\hat{\mathbf{k}}) \hat{g}(\mathbf{k})\right] = -\mathbf{S} g(\mathbf{r}) \tag{12}$$

where $\hat{g}(\mathbf{k})$ is the Fourier transform of $g(\mathbf{r}) = \boldsymbol{\chi}(\mathbf{r}) \rho(\mathbf{r})$ and the differential operator $\mathbf{S}$ has the following form:

$$\mathbf{S} = \frac{\vec{\nabla} \otimes \vec{\nabla}^\mathrm{T}}{\vec{\nabla}^\mathrm{T} \cdot \vec{\nabla}} - \frac{1}{3} = \frac{1}{\nabla^2} \begin{bmatrix} \partial_{xx} - \dfrac{\nabla^2}{3} & \partial_{xy} & \partial_{xz} \\ \partial_{yx} & \partial_{yy} - \dfrac{\nabla^2}{3} & \partial_{yz} \\ \partial_{zx} & \partial_{zy} & \partial_{zz} - \dfrac{\nabla^2}{3} \end{bmatrix} \tag{13}$$

This is a reciprocal equation to Eq. (3) – multiplication by a vector in Fourier space is equivalent to taking a gradient in real space and division by $r^2$ in real space is equivalent to the inverse Laplacian in *k*-space.

In the case where both the probability density and the magnetic susceptibility tensor are position-dependent, this operator acts on their product and the following general expression is obtained for the matrix elements of the paramagnetic shift tensor:

$$\sigma_{ij}(\mathbf{r}) = -\frac{1}{\nabla^2} \sum_k \partial_{ik} \left[\chi_{kj}(\mathbf{r}) \rho(\mathbf{r})\right] + \frac{\delta_{ij}}{3} \chi_{ij}(\mathbf{r}) \rho(\mathbf{r}) \tag{14}$$

The second term in the right hand side effectively subtracts the Fermi contact part of the full paramagnetic shift, ensuring that PCS does not depend on $\text{Tr}(\chi)$ in the same way as it happens in the point model.

Eq. (14) simplifies significantly under the assumption that the susceptibility tensor is the same at all locations. After the isotropic average is taken, the result is:

$$\sigma(\mathbf{r}) = -\frac{1}{3}\left[\frac{\vec{\nabla}^T \cdot \chi \cdot \vec{\nabla}}{\vec{\nabla}^T \cdot \vec{\nabla}} - \frac{1}{3}\text{Tr}(\chi)\right]\rho(\mathbf{r}) \tag{15}$$

In the Fourier space this equation can be written as

$$\hat{\sigma}(\mathbf{k}) = -\frac{1}{3}\left[\frac{\mathbf{k}^T \cdot \chi \cdot \mathbf{k}}{\mathbf{k} \cdot \mathbf{k}^T} - \frac{1}{3}\text{Tr}(\chi)\right]\hat{\rho}(\mathbf{k}) \tag{16}$$

where $\hat{\rho}(\mathbf{k})$ is the Fourier transform of $\rho(\mathbf{r})$ and $\hat{\sigma}(\mathbf{k})$ is the Fourier transform of $\sigma(\mathbf{r})$.

In the derivation presented above, $\rho(\mathbf{r})$ is the *statistical* probability density of the paramagnetic centre, but one can make an approximate parallel here with the *quantum mechanical* spin density. The dipolar part of the hyperfine coupling at the nuclear position $\mathbf{r}$ for a given spin density $\rho^{\text{spin}}(\mathbf{r}_e)$ is

$$\mathbf{A}^{\text{dip}}(\mathbf{r}) = \frac{3\mu_0 \gamma_e \gamma_n \hbar}{4\pi}\int \frac{\mathbf{D}(\mathbf{r}_e - \mathbf{r})}{|\mathbf{r}_e - \mathbf{r}|^3}\rho^{\text{spin}}(\mathbf{r}_e)\,d^3\mathbf{r}_e \tag{17}$$

and PCS computed *ab initio* leads to the same equation but with the spin density

$$\sigma(\mathbf{r}) = -\frac{1}{3}\frac{\text{Tr}\left[\mathbf{A}^{\text{dip}}(\mathbf{r}) \cdot \chi\right]}{\hbar \mu_0 \gamma_e \gamma_n}, \tag{18}$$

meaning that the dipolar hyperfine coupling tensor field, viewed as an integral over the spin density, can also be expressed using the differential operator from Eq. (13):

$$\mathbf{A}^{\text{dip}}(\mathbf{r}) = -\hbar \mu_0 \gamma_e \gamma_n \mathbf{S}\rho^{\text{spin}}(\mathbf{r}). \tag{19}$$

It must be noted that Eq. (18) only accounts for the dipolar part of the hyperfine coupling and does not include the orbital contribution, which is important in the immediate vicinity of heavy ions.[32,33]

## 4. Analytical solution to the direct problem

The "direct" problem will be defined here as the task of calculating PCS from a given magnetic susceptibility tensor, and a given probability distribution of the paramagnetic centre at the specified nuclear coordinates. General case with a position-dependent magnetic susceptibility tensor is described by Eq. (14). It is clear from the form of Eq. (14) that in order to disentangle effects of two spatial functions of density and susceptibility we need to know *a priory* at least one of those. Below we analyse the special case where magnetic susceptibility tensor is the same at every point of the probability density as described by Eq. (15). This approximation is reasonable for modelling mobility of lanthanide tags as it was shown by Shishmarev and Otting with "two-hinged" approximation where orientation of susceptibility tensor is the same for each rotamer.[29]

### 4.1 General solution

The easiest way to solve Eq. (15) analytically is to expand the probability density in spherical harmonics:

$$\rho(\mathbf{r}) = \sum_{l,m} \alpha_l^m(r) Y_l^m(\hat{\mathbf{r}}) \tag{20}$$

where $\alpha_l^m(r)$ are radial functions serving as expansion coefficients in this angular function series:

$$\alpha_l^m(r) = \int_\Omega Y_l^{m*}(\hat{\mathbf{r}}) \rho(\mathbf{r}) d^2\hat{\mathbf{r}} \tag{21}$$

For a spherically isotropic probability density distribution, the sum in Eq. (20) only has one term with $l=0$ and $m=0$; for a spherically anisotropic density there would also be higher terms in the spherical harmonics expansion. The Fourier transform of the density in Eq. (20) leaves the angular part the same but the radial part is integrated with spherical Bessel functions of the first kind:

$$\hat{\rho}(\mathbf{k}) = \sqrt{\frac{2}{\pi}} \sum_{l,m} (-i)^l Y_l^m(\hat{\mathbf{k}}) \int_0^\infty \alpha_l^m(s) j_l(ks) s^2 ds \tag{22}$$

We shall substitute the density from Eq. (16) into Eq. (22) and expand the products of spherical harmonics using Clebsch-Gordan coefficients. The result has three terms when $l \geq 2$ and two when $l < 2$:

$$Y_l^m Y_2^{m'} = \sqrt{\frac{5(2l+1)}{4\pi}} \left( \frac{C_{l,0,2,0}^{l-2,0} C_{l,m,2,m'}^{l-2,m+m'}}{\sqrt{2l-3}} Y_{l-2}^{m+m'} + \frac{C_{l,0,l',0}^{l,0} C_{l,m,l',m'}^{l,m+m'}}{\sqrt{2l+1}} Y_l^{m+m'} + \frac{C_{l,0,l',0}^{l+2,0} C_{l,m,2,m'}^{l+2,m+m'}}{\sqrt{2l+5}} Y_{l+2}^{m+m'} \right) \tag{23}$$

With this substitution in place, Eq. (16) acquires the following form:

$$\hat{\sigma}(\mathbf{k}) = -\frac{\sqrt{10}}{6\pi} \sum_{l,m} (-i)^l \sqrt{2l+1} \left[ \int_0^\infty j_l(ks) \alpha_l^m(s) s^2 ds \right] \sum_{m'=-2}^{2} \chi_{m'} \left[ \begin{array}{l} \dfrac{C_{l,0,2,0}^{l-2,0} C_{l,m,2,m'}^{l-2,m+m'}}{\sqrt{2l-3}} Y_{l-2}^{m+m'}(\hat{\mathbf{k}}) + \\ \dfrac{C_{l,0,2,0}^{l,0} C_{l,m,2,m'}^{l,m+m'}}{\sqrt{2l+1}} Y_l^{m+m'}(\hat{\mathbf{k}}) + \\ \dfrac{C_{l,0,2,0}^{l+2,0} C_{l,m,2,m'}^{l+2,m+m'}}{\sqrt{2l+5}} Y_{l+2}^{m+m'}(\hat{\mathbf{k}}) \end{array} \right] \tag{24}$$

To get the final answer we must take the inverse Fourier transform of Eq. (24). The spherical harmonics again remain the same, and the radial part is integrated with spherical Bessel functions. The double integrals that make an appearance are all straightforward. Changing the order of integration and integrating the product of two spherical Bessel function with respect to $k$ gives:

$$\int_0^\infty \int_0^\infty \alpha_l^m(s) s^2 j_l(ks) j_{l-2}(kr) k^2 ds dk = \pi \left(l - \frac{1}{2}\right) r^{l-2} \int_r^\infty \alpha_l^m(s) s^{-l+1} ds \tag{25}$$

$$\int_0^\infty \int_0^\infty \alpha_l^m(s) s^2 j_l(ks) j_l(kr) k^2 ds dk = \frac{\pi}{2r^2} \int_0^\infty \alpha_l^m(s) s^2 \delta(r-s) ds = \frac{\pi}{2} \alpha_l^m(r) \tag{26}$$

$$\int_0^\infty \int_0^\infty \alpha_l^m(s) s^2 j_l(ks) j_{l+2}(kr) k^2 ds dk = \pi \left(l + \frac{3}{2}\right) r^{-l-3} \int_0^r \alpha_l^m(s) s^{l+2} ds \tag{27}$$

Taking everything together we obtain the general analytical solution for Eq. (15):

$$\sigma(\mathbf{r}) = -\frac{\sqrt{20}}{12\sqrt{\pi}} \sum_{l,m} \sqrt{2l+1} \sum_{m'=-2}^{2} \chi_{m'} \left( P_{l,m,m'}^{\text{OUT}}(\mathbf{r}) + P_{l,m,m'}^0(\mathbf{r}) + P_{l,m,m'}^{\text{IN}}(\mathbf{r}) \right) \tag{28}$$

There are three physically different contributions associated with the three integrals of the radial probability density functions in Eqs. (25)-(27). The first one corresponds to the integral of the density weighted with a monotonically decreasing function over the region outside the sphere of radius $r$:

$$P^{\text{OUT}}_{l,m,m'}(\mathbf{r}) = -\frac{2l-1}{\sqrt{2l-3}} C^{l-2,0}_{l,0,2,0} C^{l-2,m+m'}_{l,m,2,m'} Y^{m+m'}_{l-2}(\hat{\mathbf{r}}) r^{l-2} \int_r^\infty \alpha_l^m(s) s^{-l+1} ds \quad (29)$$

For any nucleus positioned outside the bounding sphere of the paramagnetic centre probability density this contribution is zero. The second contribution is proportional to the probability density at the nucleus itself:

$$P^0_{l,m,m'}(\mathbf{r}) = \frac{1}{\sqrt{2l+1}} C^{l,0}_{l,0,2,0} C^{l,m+m'}_{l,m,2,m'} Y^{m+m'}_{l}(\hat{\mathbf{r}}) \alpha_l^m(r) \quad (30)$$

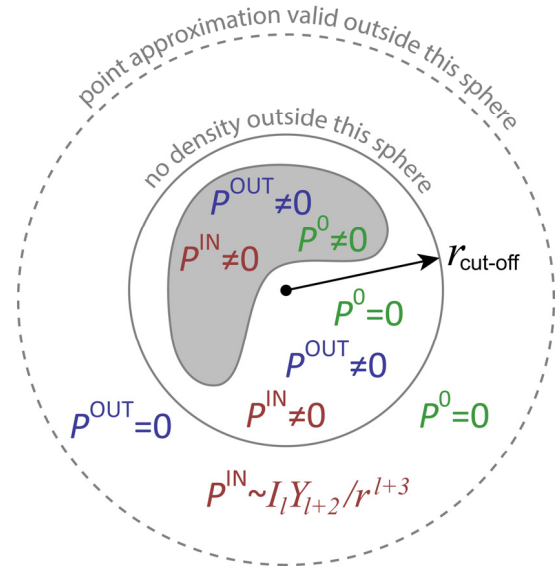

**Figure 1.** A schematic representation of the paramagnetic centre probability density (grey shaded area) and regions where different terms in Equation (28) contribute to the resulting pseudocontact shift. Outside the bounding sphere (grey solid line) only $P^{\text{IN}}$ is non-zero; inside the bounding sphere an additional term $P^{\text{OUT}}$ becomes important; $P^0$ contributes only inside the density. At a sufficient distance from the bounding sphere (grey dashed line) the point paramagnetic centre approximation becomes valid.

We are in practice unlikely to be able to measure chemical shifts of the nuclei that are directly underneath the spin density due to their fast relaxation. We are thus left with the third and the most important contribution that is associated with the part of the density that is inside the sphere of radius $r$:

$$P^{\text{IN}}_{l,m,m'}(\mathbf{r}) = -\frac{2l+3}{\sqrt{2l+5}} C^{l+2,0}_{l,0,2,0} C^{l+2,m+m'}_{l,m,2,m'} Y^{m+m'}_{l+2}(\hat{\mathbf{r}}) r^{-l-3} \int_0^r \alpha_l^m(s) s^{l+2} ds \quad (31)$$

If we assume that there is no paramagnetic centre density outside that sphere, this would allow us to extend the upper integration limit to infinity and the integrals then correspond to the multipole moments of the probability density of the paramagnetic centre:

$$I_l^m = \int_0^\infty \left[ \int_\Omega Y_l^{m*}(\hat{s}) \rho(\mathbf{s}) d^2\hat{s} \right] s^{l+2} ds \quad (32)$$

A schematic diagram of the entire argument is given in Figure 1. Outside the bounding sphere the final expression for PCS is:

$$\sigma(\mathbf{r}) = \frac{\sqrt{20}}{12\sqrt{\pi}} \sum_{l,m} \frac{1}{r^{l+3}} \frac{\sqrt{2l+1}(2l+3)}{\sqrt{2l+5}} I_l^m C_{l,0,2,0}^{l+2,0} \sum_{m'=-2}^{2} \chi_{m'} C_{l,m,2,m'}^{l+2,m+m'} Y_{l+2}^{m+m'}(\hat{\mathbf{r}}) \quad (33)$$

This solution tells us that PCS is sensitive to the multipole moments of the paramagnetic centre probability density distribution. In the case of isotropic distribution, the PCS is the same as from the point source – this follows from Eq. (33) when we put $l$ to zero; it simplifies into Eq. (7). Therefore, only the anisotropy of the density makes a difference in the PCS compared to the point source. We shall therefore proceed to explore some simple anisotropic probability densities.

More detailed derivation of Eq. (28) and Eq. (33) is provided in Supporting Information.

### 4.2 PCS from a Gaussian paramagnetic centre distribution

This section contains an analysis of the consequences of simple deviations from the point paramagnetic centre approximation. To get a quantitative idea about how far the point dipole approximation can in practice be stretched, we shall consider a family of Gaussian paramagnetic centre distributions.

It follows from the treatment above that PCS field outside any isotropic paramagnetic centre distribution is identical to the PCS field generated by a point centre. The difference form the point PCS appears only inside the density. For example, in the case of an isotropic Gaussian, where sum in Eq. (20) has only one element with $l=0$ and $m=0$, we have

$$\alpha_0^0(r) = \frac{1}{a^3 \pi \sqrt{2}} e^{-\frac{r^2}{2a^2}} \quad (34)$$

and the exact PCS is:

$$\sigma(\mathbf{r}) = \frac{\sqrt{2}}{24\pi\sqrt{\pi}r^3 a^3}\left(3\sqrt{2\pi}a^3 \text{erf}\left(\frac{r}{\sqrt{2}a}\right) - 2re^{-\frac{r^2}{2a^2}}\left(r^2 + 3a^2\right)\right) \sum_{m=-2}^{2} \chi_m Y_2^m(\hat{\mathbf{r}}) \quad (35)$$

where $a$ is the standard deviation of the Gaussian. Figure 2 demonstrates that a significant difference between the point and the isotropic Gaussian case appears only for $r < 3a$, which is not particularly interesting or dangerous because nuclei at such short distances from the paramagnetic centre are not normally visible in PCS experiments due to their rapid transverse relaxation.

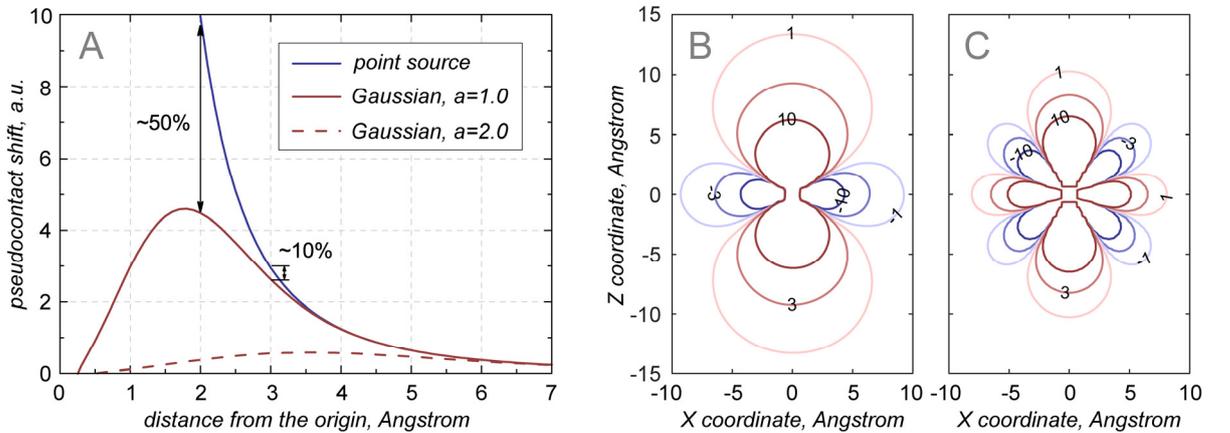

**Figure 2.** (**A**) Amplitude of the radial part of the pseudocontact shift field as a function of distance from a point source (black line) and from isotropic Gaussian sources with standard deviations of 1 Å (solid red line) and 2 Å (dashed red line). (**B, C**) PCS isolines in ppm from a prolate Gaussian source with $a = 1$ Å and $c = 3$ Å and $\chi = \text{diag}([-0.01, -0.02, 0.03])$ Å³. The Z axis of the Gaussian

is parallel to the Z axis of the susceptibility tensor eigenframe. The contribution to PCS from the isotropic part of the density is on the left and the contribution from the correction from anisotropic part is on the right.

A rather more worrying situation emerges when we consider a realistically anisotropic Gaussian distribution of the paramagnetic centre:

$$\rho(x,y,z) = \frac{1}{a^2 c \sqrt{(2\pi)^3}} \exp\left[-\frac{1}{2}\left(\frac{x^2+y^2}{a^2} + \frac{z^2}{c^2}\right)\right] \quad (36)$$

The case with $a > c$ corresponds to an oblate ellipsoid and the case with $a < c$ to a prolate one. The anisotropy results in the emergence of the next multipole in Eq. (32)

$$I_2^0 = \sqrt{5}(c^2 - a^2)/(2\sqrt{\pi}) \quad (37)$$

with the amplitude that depends on the difference of the squares of axial and equatorial sizes of the Gaussian. The resulting correction to the PCS

$$\sigma_2^0(\mathbf{r}) = \frac{\sqrt{5}(c^2 - a^2)}{24\pi} \frac{1}{r^5} \sum_{m=-2}^{2} \chi_m \sqrt{(m^2-4^2)(m^2-3^2)} Y_4^m(\hat{r}) \quad (38)$$

involves fourth rank spherical harmonics and the fifth power of the distance. It is illustrated graphically in Figure 2 – it is clear that the correction to the point dipole solution is larger than 1 ppm up to 10 Å away from the paramagnetic label.

The magnitudes and the asymptotic behaviour of the three contributions to the PCS in Eq. (28) can be summarised into the following rules:

1. For any nucleus located outside the sphere that is three times the radius of the bounding sphere of the paramagnetic centre probability density, the point paramagnetic centre approximation is valid.

2. For any nucleus located in the immediate vicinity of the bounding sphere, extra multipoles would appear in the PCS, and Eq. (33) must be used.

3. For any nucleus located inside the bounding sphere, no simplifications are available, and the full analytical solution in Eq. (28) must be used.

## 5. Analytical solution to the inverse problem

Eq. (33) is linear with respect to the components of the susceptibility tensor and the multipole moments of the probability density. Both sets of parameters are therefore easy to extract by fitting a sufficiently large data set comprising nuclear coordinates and pseudocontact shifts. However, because experimental PCS measurements inside the bounding sphere of the paramagnetic centre density are not usually realistic, only the multipole moments (rather than the density itself) may be extracted in a well-defined way.

An illustration for 300 nuclei randomly placed around a lump of paramagnetic centre probability density is given in Figure 3: the density was formed by four isotropic Gaussians with a standard deviation of 0.5 Å placed randomly within a 3×3×3 Å cube; nuclei were scattered within a 1 Å thick spherical layer at varying distances from the origin; the axiality and the rhombicity of the susceptibility tensor were chosen randomly from the typical range reported in the literature:[34] $\chi_{ax} = -0.45$ Å$^3$ and $\chi_{rh} = -0.1$ Å$^3$. PCS amplitudes in this system range from ±90 ppm (at 5 Å distance) to ±1 ppm at (20 Å distance).

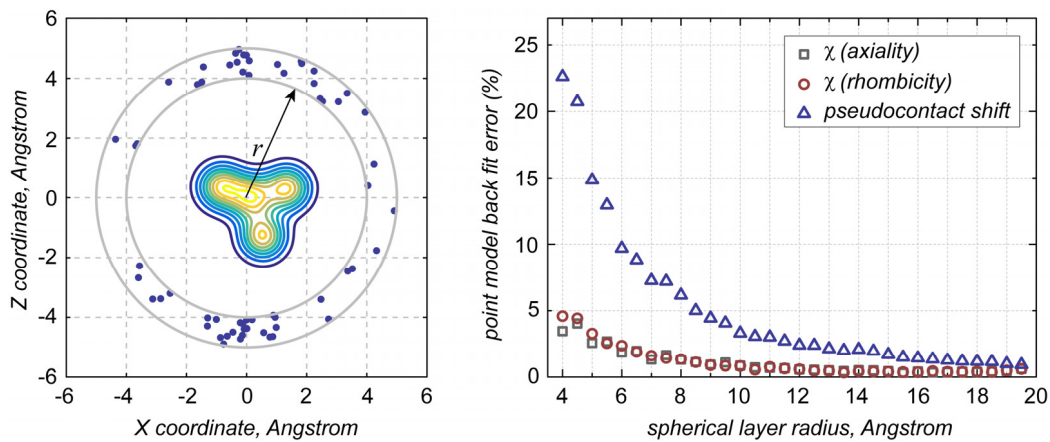

**Figure 3.** (**Left**) A schematic of the model system used to analyse the behaviour of the multipole terms in Eq. (33) – the 2D slice in the XZ plane shows the contour lines of the paramagnetic centre distribution formed by four isotropic Gaussians with a standard deviation of 0.5 Å scattered randomly within a 3×3×3 Å cube; 300 nuclei (blue dots) are scattered randomly within a 1 Å layer at varying distances from the origin. (**Right**) Point model fractional back fit error in the pseudocontact shifts and the susceptibility tensor as a function of the radius of the spherical layer in which the nuclei are scattered.

Pseudocontact shifts in this model system were calculated exactly and then an attempt was made to back-fit Eq. (33) truncated at different spherical ranks. The results were averaged over 50 instances of random nuclear position sets. As expected, the point model does not perform well below 10 Å from the bounding sphere (Figure 3, right panel). Still, even though the PCS values are badly reproduced, the susceptibility tensor is quite resilient – even at the distance of 4 Å it is recovered to within ~5% accuracy.

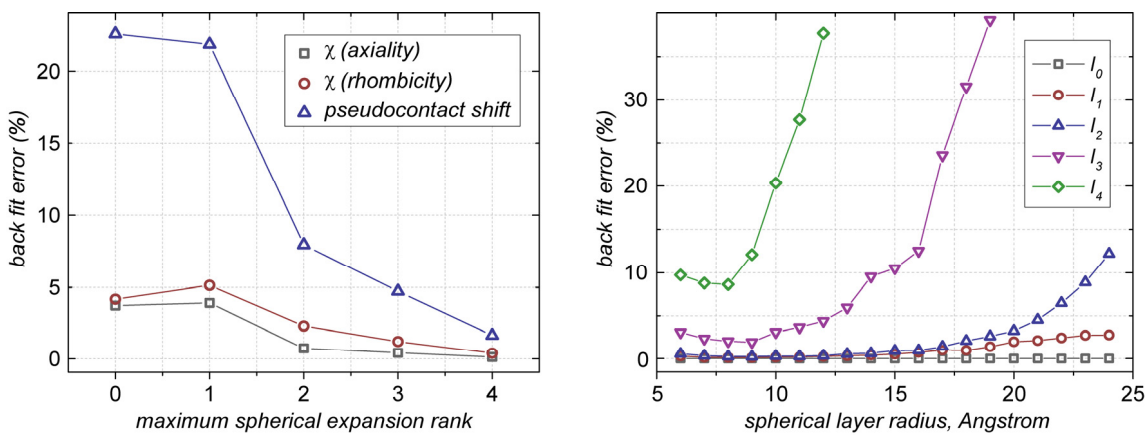

**Figure 4.** (**Left**) spherical rank dependence of the fractional back fit error produced by Eq. (33) in the case when the nuclei are scattered in the spherical layer between 4 Å and 5 Å away from the origin (a schematic is given in Figure 3). (**Right**) fractional back fit error in the multipole moments of the paramagnetic centre probability density distribution in Eq. (33) as a function of the radius of the spherical layer containing the nuclei (a schematic is given in Figure 3).

It is clear from Figure 3 that, at distances comparable to the size of the paramagnetic centre probability distribution, the point model breaks down and further terms are required in Eq. (33). Their beneficial effect is illustrated in Figure 4 – adding terms of higher spherical rank to the expansion dramatically reduces the PCS back-fit error in the vicinity of the bounding sphere.

Due to the steep distance dependence and the wobbly angle dependence of the higher multipole terms in Eq. (33), the accuracy with which these terms may be extracted also falls steeply as the nuclei are moved further away from the bounding sphere of the paramagnetic centre probability density. This is illustrated in Figure 4 – it is clear that multipole moments of spherical rank higher than 4 cannot be reliably extracted no matter how close the nuclei are to the bounding sphere. We would not therefore recommend taking Eq. (33) beyond the fourth spherical rank.

## 6. Numerical solution to the direct problem

If the probability density of the paramagnetic centre is discretised on a finite three-dimensional grid, two general numerical avenues become available for the solution of Eq. (15) – the finite-difference method and the Fourier transform method. The latter has the advantage of being fast, but the disadvantage of requiring periodic boundary conditions. This section explores both methods and comments on their relative merits.

### 6.1    Finite difference methods

A general algorithm for the generation of elementary finite difference operators of a given derivative and accuracy order on a given finite grid has been published by Fornberg.[35] For a rectangular grid with $N$ points in X direction, $M$ points in Y direction and $K$ points in Z direction, finite difference matrix representations of the relevant derivative operators acting on the vectorisation of the spin density cube are:

$$\left[\frac{\partial}{\partial x}\right] = \mathbf{D}_N^{(1)} \otimes \mathbf{1}_M \otimes \mathbf{1}_K, \quad \left[\frac{\partial^2}{\partial x^2}\right] = \mathbf{D}_N^{(2)} \otimes \mathbf{1}_M \otimes \mathbf{1}_K, \quad \left[\frac{\partial^2}{\partial x \partial y}\right] = \mathbf{D}_N^{(1)} \otimes \mathbf{D}_M^{(1)} \otimes \mathbf{1}_K \quad (39)$$

and similarly for the other first and second derivatives. In these expressions, $\mathbf{D}_N^{(k)}$ is a matrix representation of the $k$-th derivative operator on a grid with $N$ points and $\mathbf{1}_M$ is a unit matrix of dimension $M$. The dimension of the matrices in Eq. (39) is $NMK$. For a typical grid with 256 points in each dimension this is a large number, but because finite difference operators are local, the matrices produced by Eq. (39) are very sparse.

The matrices required for the solution of Eq. (15) are $\mathbf{K} = -\left[\vec{\nabla}^{\mathrm{T}} \cdot \boldsymbol{\chi} \cdot \vec{\nabla}\right]/3$ and $\mathbf{L} = \left[\vec{\nabla}^{\mathrm{T}} \cdot \vec{\nabla}\right]$. In terms of these matrices, the solution may be written as

$$\boldsymbol{\sigma} = \mathbf{L}^{-1}\mathbf{K}\boldsymbol{\rho} \quad (40)$$

where $\boldsymbol{\rho}$ is the vectorization of the probability density cube on the chosen grid and $\boldsymbol{\sigma}$ is the vectorisation of the pseudocontact shift field on the same grid; here and below we take only the traceless part of the $\boldsymbol{\chi}$ tensor. The final step is to project out the PCS values on the nuclei, for which the interpolation matrix $\mathbf{P}_{\mathrm{N}}$, also published by Fornberg in the same paper,[35] may be used. The final expression is

$$\boldsymbol{\sigma}_{\mathrm{N}} = \mathbf{P}_{\mathrm{N}}\mathbf{L}^{-1}\mathbf{K}\boldsymbol{\rho} \quad (41)$$

where the action by the very sparse inverse Laplacian matrix on the $\mathbf{K}\boldsymbol{\rho}$ vector is beset computed using an iterative solver, such as GMRES.[36] Here and below the $\mathbf{L}^{-1}$ symbol should be understood in that sense – the inverse Laplacian is never computed explicitly.

This method is very easy to set up – see, for example, the `kpcs.m` function supplied with Spinach library.[37] Its downside is unfavourable scaling: the dimension of the finite difference matrices involved is $KMN$ and the scaling of sparse solvers is approximately quadratic in the matrix dimension, meaning that the numerical complexity of this method scales approximately as the sixth power of the grid size. On a contemporary computer, a solution for a 256×256×256 point grid takes about an hour.

### 6.2    Fourier transform methods

An alternative method for solving Eq.(15) takes advantage of the Fourier domain expression in Eq. (16), where Fourier transform can be taken numerically using the fast Fourier transform (FFT) algorithm.[38] Using the same interpolation operator as in Eq. (41), we get

$$\boldsymbol{\sigma}_N = -\frac{1}{3}\mathbf{P}_N \cdot \mathrm{Re}\left[\mathrm{FFT}_-\left\{\frac{\mathbf{k}^T \cdot \boldsymbol{\chi} \cdot \mathbf{k}}{\mathbf{k}^T \cdot \mathbf{k}}\mathrm{FFT}_+\left\{\rho(\mathbf{r})\right\}\right\}\right] \qquad (42)$$

where $\chi$ stands for the traceless part of the magnetic susceptibility tensor. This method is about the same as the finite difference method in terms of the implementation complexity, but it is much faster – the complexity scaling of the three-dimensional fast Fourier transform is $NMK\log(N)\log(M)\log(K)$, which is close to cubic scaling with respect to the grid size. This takes the simulation time into the region of seconds. This method is particularly fast on modern computing hardware because fast Fourier transform routines are available for GPGPU coprocessor cards (*NVidia Tesla K40* cards were used in this work). The performance gain is shown in the left panel of Figure 5.

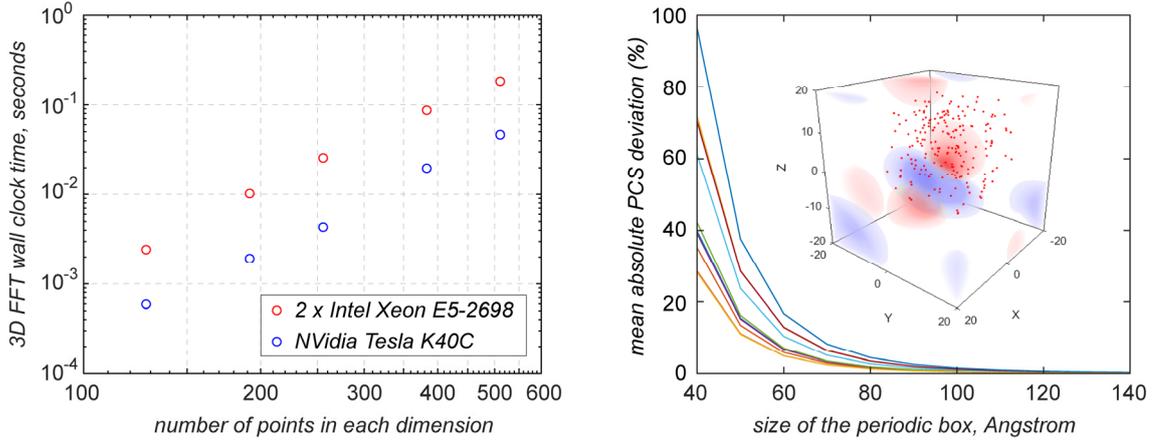

**Figure 5.** (**Left**) 3D FFT wall clock time as a function of the grid size for the calculation performed on the CPU *vs.* a dedicated coprocessor card. (**Right**) PCS error due to the periodic boundary conditions in the FFT method as a function of the periodic box size. The nuclei (red points in the inset) are placed randomly in the 20 Å box in positive octant; the paramagnetic centre has an isotropic Gaussian distribution centred at [–10, –10, –10] Å with a standard deviation of 1.0 Å.

The only significant problem with the FFT method is the periodic boundary condition – care must be taken to ensure that the images do not contribute significantly to the solution. Thankfully, PCS decays cubically with distance; it is sufficient for the distance from the paramagnetic centre to the cube boundary and the distance from the nucleus to the cube boundary to be about three times larger than the distance from the paramagnetic centre to the nucleus – a numerical example is given in Figure 5.

## 7. Numerical solution to the inverse problem

The "inverse problem" refers to the task of reconstructing the paramagnetic centre probability density from the experimental values of pseudocontact shifts and atomic coordinates. The task may be formulated as finding the probability density $\rho(\mathbf{r})$ that minimises the following functional:

$$\Omega[\rho(\mathbf{r})] = E[\rho(\mathbf{r})] + T[\rho(\mathbf{r})], \qquad (43)$$

where the least squares error is

$$E[\rho(\mathbf{r})] = \left\|\mathbf{P}_N \mathbf{L}^{-1}\mathbf{K}\rho(\mathbf{r}) - \boldsymbol{\sigma}_{\mathrm{expt}}\right\|^2 \qquad (44)$$

and the standard Tikhonov regularisation term emphasizing smooth solutions is[39]

$$T[\rho(\mathbf{r})] = \lambda\left\|\mathbf{L}\rho(\mathbf{r})\right\|^2, \qquad (45)$$

where $\lambda$ is a user-specified regularisation parameter. Regularisation is only needed when the inverse problem is ill-posed, *i.e.* the number of points in the probability density grid exceeds the number of experimental

PCS points and/or all experimental points are outside the bounding sphere of the density. In that case the recovered density is only an approximation that agrees with the experimental data and has the required smoothness at the same time – this point is discussed further in Section 8.

## 7.1 Finite difference method

In order to proceed with minimising $\Omega[\rho(\mathbf{r})]$ with respect to $\rho(\mathbf{r})$, the state-of-the-art Newton-Raphson minimisers require analytical expressions for its first and second variation.[40] In a matrix representation on a discrete grid, these variations become the gradient vector and the Hessian matrix. The following vector relations are useful as a starting point.

$$\frac{\partial}{\partial \mathbf{x}^{\mathrm{T}}}\|\mathbf{A}\mathbf{x}-\mathbf{b}\|^2 = 2\mathbf{A}^{\mathrm{T}}(\mathbf{A}\mathbf{x}-\mathbf{b})$$
$$\frac{\partial^2}{\partial \mathbf{x}^{\mathrm{T}}\partial \mathbf{x}}\|\mathbf{A}\mathbf{x}-\mathbf{b}\|^2 = 2\mathbf{A}^{\mathrm{T}}\mathbf{A} \qquad (46)$$

If we use $\boldsymbol{\rho}$ to denote the vectorised form of a discrete representation of $\rho(\mathbf{r})$ on a finite three-dimensional grid, the expressions for the gradient and the Hessian of the least squares error part of Eq. (43) are:

$$\frac{\partial}{\partial \boldsymbol{\rho}^{\mathrm{T}}}\|\mathbf{PL}^{-1}\mathbf{K}\boldsymbol{\rho}-\boldsymbol{\sigma}_{\mathrm{expt}}\|^2 = 2\mathbf{KL}^{-1}\mathbf{P}^{\mathrm{T}}(\mathbf{PL}^{-1}\mathbf{K}\boldsymbol{\rho}-\boldsymbol{\sigma}_{\mathrm{expt}}) = 2\mathbf{KL}^{-1}\mathbf{P}^{\mathrm{T}}(\boldsymbol{\sigma}_{\mathrm{theo}}-\boldsymbol{\sigma}_{\mathrm{expt}})$$
$$\frac{\partial^2}{\partial \boldsymbol{\rho}^{\mathrm{T}}\partial \boldsymbol{\rho}}\|\mathbf{PL}^{-1}\mathbf{K}\boldsymbol{\rho}-\boldsymbol{\sigma}_{\mathrm{expt}}\|^2 = 2\mathbf{KL}^{-1}\mathbf{P}^{\mathrm{T}}\mathbf{PL}^{-1}\mathbf{K} \qquad (47)$$

where advantage was taken of the fact that $\mathbf{L}$ and $\mathbf{K}$ are symmetric matrices and that $\boldsymbol{\sigma}_{\mathrm{theo}} = \mathbf{PL}^{-1}\mathbf{K}\boldsymbol{\rho}$ would normally already be computed and stored in memory by the time the derivatives are requested. Although the Hessian in Eq. (47) is formally a matrix of a very large dimension, in practice only the result of its multiplication by a vector is required by the modern Newton-Raphson optimisers because they employ iterative sparse linear solvers for the inverse-Hessian-times-gradient operation.[41,42]

Similarly, the gradient and the Hessian for the Tikhonov regularisation term are:

$$\frac{\partial}{\partial \boldsymbol{\rho}^{\mathrm{T}}}T[\boldsymbol{\rho}] = 2\lambda \mathbf{L}^{\mathrm{T}}\mathbf{L}\boldsymbol{\rho}, \qquad \frac{\partial^2}{\partial \boldsymbol{\rho}^{\mathrm{T}}\partial \boldsymbol{\rho}}T[\boldsymbol{\rho}] = 2\lambda \mathbf{L}^{\mathrm{T}}\mathbf{L} \qquad (48)$$

It is in practice unnecessary to consider every point of the grid to be a variable. The location of the paramagnetic centre is usually known approximately, and it is sufficient to only consider the points in the immediate vicinity – typically a 20×20×20 Å cube around the expected location.

## 7.2 Fourier transform method

About a factor of a hundred in performance may be gained (at the cost of worrying about the periodic boundary as described in Section 6.2) by using the Fourier solution in Eq.(42). The overall error functional is still the same as in Eq. (43), but the terms for the least squares error and the Tikhonov regularisation now involve three-dimensional fast Fourier transforms:

$$E[\rho(\mathbf{r})] = \|\boldsymbol{\sigma}_{\mathrm{theo}}-\boldsymbol{\sigma}_{\mathrm{expt}}\|^2, \qquad T[\rho(\mathbf{r})] = \lambda \operatorname{Re}\left[\mathrm{FFT}_{-}\left\{\mathbf{k}\cdot\mathbf{k}^{\mathrm{T}}\mathrm{FFT}_{+}\left\{\rho(\mathbf{r})\right\}\right\}\right]^2 \qquad (49)$$

The first variation and the action by the second variation of $E[\rho(\mathbf{r})]$ on a probe function $\eta(\mathbf{r})$ are:

$$\frac{\delta E}{\delta \rho} = -\frac{2}{3} \text{Re}\left[ \text{FFT}_{-} \left\{ \frac{\mathbf{k}^{T} \cdot \boldsymbol{\chi} \cdot \mathbf{k}}{\mathbf{k} \cdot \mathbf{k}^{T}} \text{FFT}_{+} \left\{ \mathbf{P}_{N}^{T} \left[ \boldsymbol{\sigma}_{\text{theo}} - \boldsymbol{\sigma}_{\text{expt}} \right] \right\} \right\} \right]$$

$$\frac{\delta^{2} E}{\delta \rho^{2}} \left[ \eta(\mathbf{r}) \right] = -\frac{2}{3} \text{Re}\left[ \text{FFT}_{-} \left\{ \frac{\mathbf{k}^{T} \cdot \boldsymbol{\chi} \cdot \mathbf{k}}{\mathbf{k} \cdot \mathbf{k}^{T}} \text{FFT}_{+} \left\{ \mathbf{P}_{N}^{T} \mathbf{P}_{N} \text{FFT}_{-} \left\{ \frac{\mathbf{k}^{T} \cdot \boldsymbol{\chi} \cdot \mathbf{k}}{\mathbf{k} \cdot \mathbf{k}^{T}} \text{FFT}_{+} \left\{ \eta(\mathbf{r}) \right\} \right\} \right\} \right\} \right]$$

(50)

The corresponding variations for the Tikhonov regularisation functional are:

$$\frac{\delta T}{\delta \rho} = 2\lambda \text{Re}\left[ \text{FFT}_{-} \left\{ \left( \mathbf{k} \cdot \mathbf{k}^{T} \right)^{2} \text{FFT}_{+} \left\{ \rho(\mathbf{r}) \right\} \right\} \right]$$

$$\frac{\delta^{2} T}{\delta \rho^{2}} \left[ \eta(\mathbf{r}) \right] = 2\lambda \text{Re}\left[ \text{FFT}_{-} \left\{ \left( \mathbf{k} \cdot \mathbf{k}^{T} \right)^{2} \text{FFT}_{+} \left\{ \eta(\mathbf{r}) \right\} \right\} \right]$$

(51)

A numerical technicality in Eqs (49) and (50) concerns the behaviour of $\mathbf{k}^{T} \cdot \boldsymbol{\chi} \cdot \mathbf{k} / \mathbf{k}^{T} \cdot \mathbf{k}$ term around the origin of the *k*-space. It is easy to demonstrate that the limit value

$$\lim_{|\mathbf{k}| \to 0} \frac{\mathbf{k}^{T} \cdot \boldsymbol{\chi} \cdot \mathbf{k}}{\mathbf{k} \cdot \mathbf{k}^{T}}$$

(52)

corresponds to the total integral of the solution in the real space. In the case of pseudocontact shift this integral is known to be zero; this resolves the ambiguity.

### 7.3 Regularisation parameters

The standard way of choosing the Tikhonov regularisation parameter is the L-curve method:[43,44] the error functional minimisation is repeated with different values of $\lambda$ and the value is picked that corresponds to the maximum curvature $\kappa(\lambda)$ of the following curve

$$\begin{cases} x(\lambda) = \ln\left( E_{\lambda}\left[ \boldsymbol{\rho}_{\text{opt}} \right] \right) \\ y(\lambda) = \ln\left( T_{\lambda}\left[ \boldsymbol{\rho}_{\text{opt}} \right] \right) \end{cases} \qquad \kappa(\lambda) = \frac{|x'y'' - x''y'|}{\left( x'^{2} + y'^{2} \right)^{3/2}}$$

(53)

In practice a discrete point set $\{E_{\lambda_k}, T_{\lambda_k}\}$ for different regularisation parameter values $\lambda_k$ is interpolated by a fifth order spline and then differentiated numerically. The standard third order spline is not sufficient here because second derivatives are involved in the definition of curvature that later needs to be differentiated again to find a maximum. The derivatives of the spline are fed into the expression for the curvature and the maximum is found with respect to $\lambda$ using standard optimisation techniques.

## 8. Probability density reconstruction examples

Paramagnetic centre probability density reconstruction and its interpretation for real-life paramagnetic protein systems is a very large block of work that has been submitted for publication as a separate paper. This communication deals with the technical side of the matter; here we shall therefore only look at two basic example cases and make some general cautionary comments.

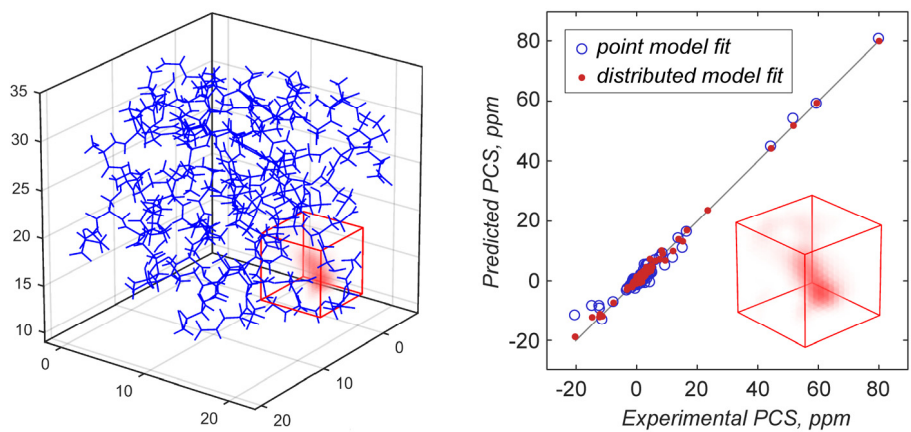

**Figure 6.** Paramagnetic centre probability density reconstruction for $Tm^{3+}$ labelled calbindin $D_{9K}$ (PDB code: 1IGV),[34,45] for which the point approximation is known to produce a poor fit (right panel, blue circles). Numerical probability density reconstruction (left panel) reveals that the thulium ion probability density has an elongated shape, likely as a consequence of local conformational mobility or multiple competing oxygen coordination sites. The red cube is the volume in which the probability density is allowed to vary during the Tikhonov regularised reconstruction process.

An example, shown in Figure 6, deals with the paramagnetic centre density distribution in the well characterized calbindin $D_{9K}$ in which one of the two calcium ions has been replaced by a thulium ion.[45] Detailed coordinate and PCS data for this protein is available in the supplementary information of the paper by the Florence group.[34] The point paramagnetic centre model produces a rather unsatisfactory fit in the vicinity of the metal (Figure 6, right panel). A reasonable explanation (proposed to the authors by Gottfried Otting and Thomas Huber) would be that fast exchange between metal binding sites could be present. We can now confirm that there is indeed an elongated distribution in the metal position (Figure 6, left panel).

Simple and intuitive though the reconstructions of the kind shown in Figure 6 might appear, a few words of caution are in order. An important consequence of the asymptotic behaviour of the three components of the exact solution in Eq. (28) is that the fine *internal* details of the paramagnetic centre probability density distribution are expected to be lost if the PCS data is measured for the nuclei that are positioned *outside* the bounding sphere of the density. In practice this means that the shape of the bounding surface of the probability distribution is well reproduced by the reconstruction, but its internal details are not.

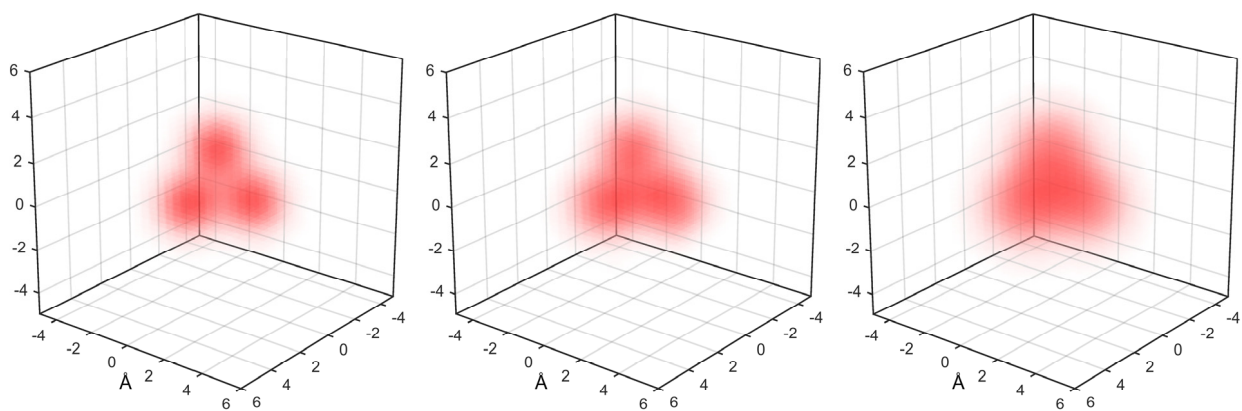

**Figure 7.** *Left panel*: a trimodal paramagnetic centre probability density distribution composed of three isotropic Gaussians with a standard deviation of 1 Å placed at [2, 0, 0], [0, 2, 0] and [0, 0, 2] Angstrom. *Middle panel*: probability density reconstruction from the PCS values computed at 100 nuclei placed randomly throughout a 10×10×10 Å cube centred at the origin. *Right panel*: probability density reconstruction from the PCS computed at 500 nuclei placed inside the same 10×10×10 Å cube, but outside the 5 Å sphere centred at density.

The consequences are illustrated in Figure 7. The left diagram shows a simple test density composed of three Gaussian functions. The middle diagram shows the reconstruction performed using PCS data from the nuclei that are distributed randomly within a 10 Å cube around the origin, including inside the density. The right diagram shows the reconstruction performed using PCS data from the nuclei that are placed outside the 5 Å sphere around the density. The latter case is what typically happens in paramagnetic NMR; it is clear that a reasonable likeness of the true probability density is recovered, but the internal details of the distribution are lost because PCS values at those points are not sensitive to the radial functions in the probability density expansion in Eq. (20). This is both the consequence of the structure of the multipolar expansion and the price to pay for a Tikhonov solution to what is in general an ill-posed problem.

## 9. Conclusions

Delocalisation of the paramagnetic centre creates deviations from the point dipole model for the pseudocontact shift. The analytical solution to the partial differential equation for PCS indicates that the key factor in such deviations is the angular anisotropy in the probability density – PCS from spherical paramagnetic centre distributions is identical to the point dipole PCS outside the bounding sphere of the density.

Multipoles with spherical rank $l$ in the paramagnetic centre distribution create rank $l+2$ multipoles in the resulting PCS field that decay with the distance as $1/r^{l+3}$ outside the bounding sphere of the density. Due to fast decay of these higher rank components, only terms up to spherical rank four can in practice be extracted by fitting experimental data. The behaviour of the PCS field inside the bounding sphere is more complicated and may be analysed numerically; however, experimental measurements in such close proximity to the paramagnetic centre are rarely possible.

Discretising the paramagnetic centre probability density and the PCS field on finite grids makes it possible to solve the partial differential equation for PCS numerically, using either finite difference operators or fast Fourier transforms. The ill-posed inverse problem of reconstructing the probability density from experimental PCS data can then be solved using Tikhonov regularisation under the assumption that the paramagnetic centre has the same effective magnetic susceptibility anisotropy at every point in the distribution.

The methods described in this paper are implemented in versions 1.8 and later of *Spinach* library.[37]

## 10. Acknowledgments

This work was made possible by a grant from EPSRC (EP/N006895/1). Stimulating discussions with Thomas Huber, Alexander Karabanov, Claudio Luchinat, Gottfried Otting and David Parker are gratefully acknowledged.